\RequirePackage{fix-cm}
\documentclass[smallextended]{svjour3}       % onecolumn (second format)
\smartqed  % flush right qed marks, e.g. at end of proof
\usepackage{graphicx}
\usepackage{amssymb,amsmath}

\newcommand{\DETAILS}[1]{}

\newcommand{\I}{\mathrm{i}}
\newcommand\D{\mbox{d}}

\newcommand{\R}{\mathbb R}
\newcommand{\C}{\mathbb C}
\newcommand{\e}{\eqref}
\newcommand{\p}{\partial}
\newcommand{\upi}{\pi}

\newcommand \bfr{{\bf r}}

\begin{document}

\title{Poles and branch cuts in  free surface hydrodynamics%\thanks{Grants or other notes
}


\author{P. M. Lushnikov         \and
        V. E. Zakharov %etc.
}


\institute{P. M. Lushnikov \at
              Department of Mathematics and
Statistics, University of New Mexico, Albuquerque, NM 87131, USA
%              Tel.: +123-45-678910\\
%              Fax: +123-45-678910\\
           \\  Landau Institute for Theoretical Physics,
Chernogolovka, 142432, Russia \\ \email{plushnik@math.unm.edu}
       \and    V. E. Zakharov \at
              Landau Institute for Theoretical Physics,
Chernogolovka, 142432, Russia \\
Center for Advanced Studies,
Skoltech, Moscow, 143026, Russia \\
Department of Mathematics, University of Arizona, Tucson, AZ
85721, USA
}

\date{Received: date / Accepted: date}

\maketitle

\begin{abstract}
We consider the motion of ideal incompressible fluid with free surface. We analyzed the exact fluid dynamics though the time-dependent conformal mapping $z=x+iy=z(w,t)$ of the lower complex half plane of the conformal variable $w$ into the area occupied by fluid. We established the exact results on the existence vs. nonexistence of the pole and power law branch point solutions for $1/z_w$ and the complex velocity. We also proved the nonexistence of the time-dependent rational solution of that problem for the second and the first order moving pole.
\end{abstract}

\section{Introduction }
\label{sec:Introduction}

\begin{figure}
\includegraphics[width=0.99859\textwidth]{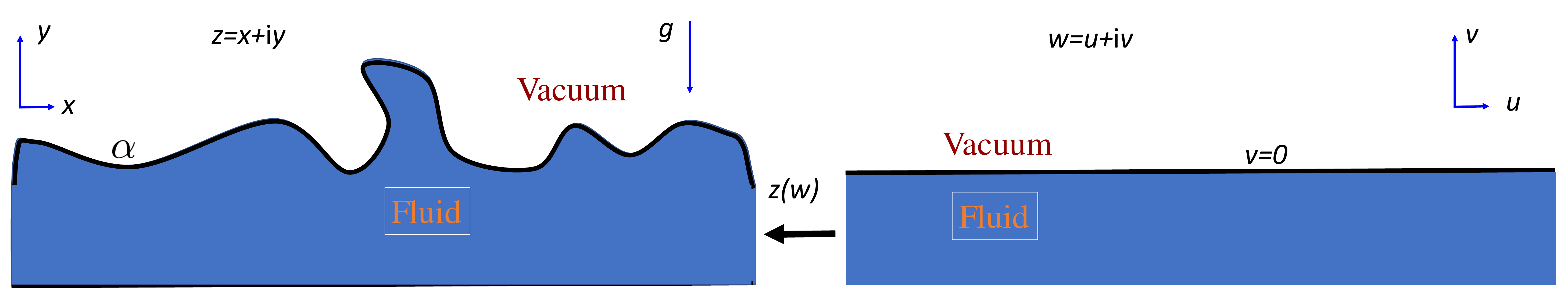}
\caption{ Shaded area represents the domain occupied by fluid in the
physical plane $z=x+\I y $ (left) and the same domain in   $w=u+\I
v$ plane  (right). Thick solid lines correspond to the fluid's free
surface.} \label{fig:schematic1}
\end{figure}

Consider  an ideal incompressible fluid with free surface which occupies the infinite region $-\infty < x < \infty$ in the
horizontal direction $x$ and extends down to $y\to -\infty$ in the
vertical direction $y$   as schematically shown on the left panel of
Fig. \ref{fig:schematic1}. It is assumed that there is no dependence on the third spatial dimension, i.e. the fluid motion is exactly two dimensional.  The bulk of fluid is at the rest, i.e. there is no motion  both at
$|x|\to \pm\infty   $ and $y\to -\infty$.  A potential motion of
the ideal incompressible fluid with free surface can be addressed by a time-dependent conformal mapping
\begin{equation} \label{zwdef}
z(w,t)=x(w,t)+\I y(w,t)
\end{equation}
of the lower complex half-plane $\mathbb{C}^-$ of the auxiliary complex variable
%
%\begin{equation} \label{wdef}
$w\equiv u+\I v, \quad -\infty<u<\infty,
$
%\end{equation}
%
into the area in $(x,y)$ plane occupied by the fluid. Here the real line $v=0$ is mapped into the fluid free surface (see Fig. \ref{fig:schematic1}) and $\mathbb{C}^-$ is defined by the condition  $-\infty<v\le0$.
 The time-dependent fluid free surface is represented in the parametric form as
\begin{equation} \label{xyu}
x=x(u,t), \ y=y(u,t).
\end{equation}
A decay of perturbation of fluid beyond flat surface at  $x(u,t)\to \pm \infty$  and/or $y\to -\infty$ requires that %
\begin{equation} \label{zlimit}
z(w,t)\to w \ \text{for} \ |w|\to\infty, \ w\in\C^-.
\end{equation}The conformal mapping \e{zwdef} implies that $z(w,t)$ is the analytic function of   $w\in\mathbb{C^-} $ and  %
\begin{equation} \label{zwconformal}
z_w\ne 0 \ \text{for any} \  w\in\mathbb{C^-}.
\end{equation}

To account for the fluid motion one  considers a complex velocity potential $\Pi(z,t)$,%
\begin{equation} \label{ComplexPotentialdef}
\Pi=\Phi+\I \Theta.
\end{equation}
%
 %where %
%\begin{equation} \label{zdef}
%z=x+\I y
%\end{equation}
%
where  $\Phi(\bfr,t)$  is the
velocity potential determined by the condition that the fluid velocity ${\bf v}$    is the potential one,  ${\bf v}= \nabla \Phi,$ and $\Theta$ is the  stream  function $\Theta$ defined by%
\begin{equation} \label{Thetadef}
\Theta_x=-\Phi_y \ \text{and} \ \Theta_y=\Phi_x.
\end{equation}
The
incompressibility condition $\nabla \cdot {\bf v} = 0$ implies the
Laplace equation
%
%\begin{align} \label{laplace}
$\nabla^2 \Phi = 0$
%\end{align}
inside fluid, i.e. $\Phi$ is the harmonic function inside fluid. Eqs.
\e{Thetadef} are the Cauchy-Riemann equations ensuring the analyticity of
$\Pi(z,t)$ in the domain of $z$ plane occupied by the fluid so $\Theta$ is
the harmonic  conjugate of $\Phi.$ Without loss of generality, we assume a
zero  Dirichlet boundary condition (BC) for $\Pi$ as

\begin{equation} \label{Dirichlet2}
\Pi\to 0 \  \text{} \ \text{for } \ |x|\to \infty \ \text{or} \ y\to - \infty.
\end{equation}

  The conformal mapping \e{zwdef} ensures that the function $\Pi(z,t)$~ %
\e{ComplexPotentialdef} transforms into $\Pi(w,t)$ which is the analytic function of $w$ for $w\in\mathbb{C^-}$ (in the bulk of
fluid).
Here and below we abuse the notation and use the  same symbols for functions of either  $w$ or $z$     (in other words, we  assume that e.g. $\tilde \Pi(w,t)= \Pi(z(w,t),t) $ and remove $\tilde ~$ sign).
The conformal transformation    \e{zwdef} also ensures   the Cauchy-Riemann equations $
\Theta_u=-\Phi_v, \quad \Theta_v=\Phi_u $
  in $w$ plane.

BCs at the free surface  are time-dependent and consist of kinematic and dynamic BCs.
A kinematic BC ensures that  free surface moves with the normal velocity component $v_n$ of fluid particles at the free surface.   Motion of the free surface is determined by a time derivative of the parameterization  \e{xyu}  while the kinematic BC is given by a projection of $\bf v$ into the normal direction as
\begin{equation} \label{kinematicu0}
{\bf n}\cdot\left(x_t,y_t \right )=v_n\equiv{\bf n}\cdot\nabla \Phi|_{x=x(u,t),\ y=y(u,t)},
\end{equation}
where %
%\begin{equation} \label{ndef}
${\bf n}=\frac{(-y_u,x_u)}{(x_u^2+y_u^2)^{1/2}}
$%\end{equation}
is the outward unit normal vector to the free surface and subscripts here and below means partial derivatives, $x_t\equiv\frac{\p x(u,t)}{\p t}$ etc.

 A dynamic BC is given by the time-dependent Bernoulli equation (see e.g. \cite{LandauLifshitzHydrodynamics1989})
at the free surface,%
\begin{align} \label{dynamic1}
\left.\left(\Phi_t +
 \dfrac{1}{2}\left(\nabla \Phi\right)^2+gy\right)\right|_{x=x(u,t),\ y=y(u,t)}  = -P_\alpha,
\end{align}
 where $g$ is the acceleration due to gravity and %
%\begin{equation} \label{Palpha}
$P_\alpha=-\frac{\alpha(x_uy_{uu}-x_{uu}y_u) }{(x_u^2+y_u)^{3/2}}
$ %\end{equation}
  is the pressure jump  at the free
surface due to the surface tension coefficient $\alpha$. Here without loss of generality we assumed that pressure is zero above the free surface (i.e. in vacuum). Also below in this paper we assume zero surface tension $\alpha=0.$
All results below apply both to the surface gravity wave case ($g>0$) and the Rayleigh-Taylor problem $(g<0)$.

Eqs.
 \e{kinematicu0} and \e{dynamic1}  together with the analyticity (with respect to the independent variable $w$) of both $z(w,t)$ and $\Pi(w,t)$ inside fluid form a closed set of equations which is equivalent to Euler equations for dynamics of ideal fluid with free surface. The idea of using time-dependent conformal transformation like
\e{zwdef} to address free surface dynamics of ideal fluid    was exploited
by several authors
including~\cite{Ovsyannikov1973,MeisonOrzagIzraelyJCompPhys1981,TanveerProcRoySoc1991,TanveerProcRoySoc1993,DKSZ1996,ChalikovSheininAdvFluidMech1998,ChalikovSheininJCompPhys2005,ChalikovBook2016,ZakharovDyachenkoVasilievEuropJMechB2002}.
We follow the approach of Refs.
\cite{DKSZ1996,ZakharovDyachenkoArxiv2012,DyachenkoLushnikovZakharovJFM2019}
  to transform from the unknowns  $z(w,t)$ and $\Pi(w,t)$ into new equivalent  ``Dyachenko''  variables (\cite{Dyachenko2001})   %
\begin{align} \label{RVvar1}
&R=\frac{1}{z_w}, \\ \label{RVvar2}
&V=\I\frac{\p \Pi}{\p z}=\I R \Pi_w.
\end{align}
Then the dynamical equations at the real line $w=u$ take the following complex form \cite{Dyachenko2001}: \begin{align}
\frac{\partial R}{\partial t} &= \I \left(U R_u - R U_u \right), \label{Reqn4}\\
\qquad U&=\hat P^-(R\bar V+\bar RV), \quad B= \hat P^-(V\bar V), \label{UBdef4}\\
\frac{\partial V}{\partial t} &= \I \left[ U V_u - RB_u\right ]+ g(R-1), \label{Veqn4}
\end{align}
where
\begin{equation} \label{Projectordef}
\hat P^-=\frac{1}{2}(1+\I \hat H)  \quad\text{and}\quad  \hat P^+=\frac{1}{2}(1-\I \hat H)
\end{equation}
 are the projector operators of a  function $q(u)$ (defined at the real
line $w=u$) into  functions $q^+(u)$ and $q^-(u)$ analytic in $w\in\mathbb{C}^-$ and
$w\in\mathbb{C}^+$, respectively, such that %
%\begin{equation} \label{qprojectiondef}
$q=q^++q^-,
$
%\end{equation}
%
i.e. $\hat P^+(q^++q^-)=q^+$ and   $\hat P^-(q^++q^-)=q^-.$
Here we assume that $q(u)\to 0$ for $u\to \pm\infty$.
Also the bar means complex conjugation and
\begin{equation} \label{Hilbertdef}
\hat H f(u)=\frac{1}{\upi} \text{p.v.}
\int^{+\infty}_{-\infty}\frac{f(u')}{u'-u}\D u'
\end{equation}
is the Hilbert
transform with $\text{p.v.}$ meaning a Cauchy principal value of the integral.
The nonlocal operator \e{Projectordef}  appears in the dynamical equations \e{Reqn4}-\e{Veqn4} because at each given moment of time one has to find the relation between the value of $\Phi$ at the free surface and its normal derivative to evolve the free surface in the physical plane $z$ using the kinematic BC \e{kinematicu0}. Such relation is nothing more then the Dirichlet-Neumann operator   \cite{CraigSulemJCompPhys1993} which can be found in $w$ plane explicitly through the Hilbert transform  \e{Hilbertdef} (see e.g. Ref. \cite{DyachenkoLushnikovZakharovJFM2019} for more discussion on that).

It was found in Refs.
\cite{ZakharovDyachenkoArxiv2012,DyachenkoLushnikovZakharovJFM2019} that the system
 \e{Reqn4}-\e{Veqn4}   is equivalent to the non-canonical Hamiltonian equations
\begin{equation} \label{implectic2}
 {\bf Q}_t=\hat R\frac{\delta H}{\delta {\bf Q}}, \quad {\bf Q}\equiv\begin{pmatrix}y \\
\psi
\end{pmatrix}
\end{equation}
for the Hamiltonian variables  $y(u,t)$ and $\psi(u,t)\equiv\Phi(u,t)$ at the real line $w=u$,   where %
%\begin{equation} \label{Rdef}
$\hat R=\hat \Omega^{-1}=\begin{pmatrix}0 & \hat R_{12} \\
\hat R_{21} &  \hat R_{22} %\\
\end{pmatrix}
$%\end{equation}
is $2\times 2$ skew-symmetric matrix operator with the components
\begin{equation}\label{Rmatrdef}
\begin{split}
& \hat R_{11}q=0,\ \hat R_{12}q=\frac{x_u}{J}q-y_u\hat H\left ( \frac{q}{J}\right ), \\
& \hat R_{21}q=-\frac{x_u}{J}q-\frac{1}{J}\hat H\left (y_u q\right ), \quad  \hat R_{21}^\dagger=-\hat R_{12},\\
 & \hat R_{22}q=-\psi_u\hat H\left ( \frac{q}{J}\right ) -\frac{1}{J}\hat H\left (\psi_u q\right ),\quad  \hat R_{11}^\dagger=-\hat R_{11}.
\end{split}
\end{equation}
We call $\hat R=\hat \Omega^{-1}$ by the ``implectic"  operator (sometimes such type of inverse of the
symplectic operator  is also called by the co-symplectic operator, see e.g. Ref. \cite{WeinsteinJDiffGeom1983}).
Here the Hamiltonian $H$ is the total energy of fluid (kinetic plus potential energy in the gravitational field and surface tension energy) which is written in terms of the Hamiltonian variables as
\begin{equation} \label{Hpsiyonly}
H=-\frac{1}{2}\int\limits^\infty_{-\infty}\psi\hat H\psi_u\D u+\frac{g}{2}\int \limits_{-\infty}^{\infty} y^2\,(1-\hat H y_u)\D  u.
\end{equation}

Eqs. \e{implectic2} allows one to define the Poisson bracket (see Ref.
\cite{DyachenkoLushnikovZakharovJFM2019})
\begin{equation}\label{PoissonBracketsRDef}
\begin{split}
& \{F,G\}%=\sum\limits^{2}_{i,j=1}\int \limits^{\infty}_{-\infty}\D u\left %( \frac{\delta F}{\delta Q_i }\hat R_{ij}  \frac{\delta G}{\delta Q_j } %\right )
=\int \limits^{\infty}_{-\infty}\D u\left ( \frac{\delta F}{\delta y }\hat R_{12}  \frac{\delta G}{\delta \psi } +\frac{\delta F}{\delta\psi }\hat R_{21}  \frac{\delta G}{\delta y }+\frac{\delta F}{\delta \psi }\hat R_{22}  \frac{\delta G}{\delta \psi }\right )
\end{split}
\end{equation}
and rewrite Eq. \e{implectic2} in terms of  Poisson mechanics as %
\begin{equation}\label{hamiltoniancanonicalpoisson}
\begin{split}
 {\bf Q}_t=\{{\bf  Q},H\}.
 \end{split}
\end{equation}
 Thus a functional $F$ is the constant of motion of Eq. \e{hamiltoniancanonicalpoisson}
provided $\{{ F},H\}=0.$

It was found in Ref. \cite{DyachenkoDyachenkoLushnikovZakharovJFM2019} that the system
 \e{Reqn4}-\e{Veqn4}    has   an arbitrary number  of the nontrivial integrals of motion constants of motion beyond the natural integrals like the Hamiltonian $H$ \e{Hpsiyonly} and the horizonal momentum (see Ref. \cite{DyachenkoLushnikovZakharovJFM2019}).
 We many of them commuting with  with each other,
i.e.  $\{{ F},G\}=0$ for the pari of such functionals $F$\ and $G$.  It was suggested in  Ref. \cite{DyachenkoDyachenkoLushnikovZakharovJFM2019}  that the existence of such
commuting integrals of motion might be a sign of the Hamiltonian
integrability of the free surface hydrodynamics.

%\newpage

In this paper we aim to address  the complimentary question (beyond the possible Hamiltonian integrability) which is to study allowed vs. not allowed classes of solutions in the system
 \e{Reqn4}-\e{Veqn4}.  To answer that question, we consider analytical continuation of Eqs.  \e{RVvar1}-\e{Veqn4}  into the complex
plane $w\in\C.$ In particular, it amounts to straightforward replacing of $u$ by $w$ in
the integral representation of  $\hat P^+q(w)$ and $\hat P^-q(w)$ as
detailed in  Ref. \cite{DyachenkoLushnikovZakharovJFM2019}.
A complex conjugation $\bar f(w)$ of $f(w)$  in Eqs. \e{Reqn4}-\e{Veqn4} and throughout this paper is understood as applied with the assumption that $f(w)$ is the complex-valued function of the real argument $w$ even if $w$ takes the complex values so that %
\begin{equation} \label{bardef}
 \bar f(w)\equiv \overline {f(\bar{w})}.
\end{equation}That definition ensures
  the analytical continuation of $f(w)$ from
the real axis  $w=u$ into the complex plane of $w\in\mathbb{C.}$ We also notice that in Eqs.  \e{Reqn4}-\e{Veqn4} and throughout this paper we use the partial derivatives over $w$ and $u$ interchangeably by assuming the analyticity in $w.$

The  goal of this paper is to analyze the existence of  complex singularities of both $R$ and $V$ in the complex plane $w$ during the nonzero duration of time. The singularities are not allowed for  $w\in\mathbb{C}^-$ because both $z$ and $\Pi$ are analytic there (inside the fluid domain)  and the zeros of $z_w$ are also excluded for  $w\in\mathbb{C}^-$  because \e{zwdef} is the conformal map. However, the singularities are generally allowed for   $w\in\mathbb{C}^+$, i.e. outside of the fluid domain. One can trivially have any singularity (including poles, branch
points, etc.) for both $R$ and $V$ for   $w\in\mathbb{C}^+$ at the initial time
$t=0.$   The important question we analyze if there are singularities that
keep their nature in the course of evolution to at least any finite duration of time. We refer to such singularities as  ``persistent''. We found that there are severe restrictions on the existence of persistent poles of arbitrary order. These restriction are given by the following theorems which are proven below in Section \ref{sec:poles}:

 {\it Theorem} 1. Assume $R$ has the  pole of the highest order $n_{max,R}\ge 1$ and $V $ has the  pole of the highest order $n_{max,V}\ge 0$  at $z=a(t), \ a\in\C^+$   with the corresponding Laurent series

\begin{align}\label{Rpoleser}
R=\sum\limits_{j=-n_{max,R}}^{-1}{R_j}{(t)(w - a)^j}+R_{reg}, \quad n_{max,R}\ge 1
\end{align}
and
\begin{align}
V=\begin{cases}\sum\limits_{j=-n_{max,V}}^{-1}{V_j}{(t)(w - a)^j}+V_{reg} \ \text{for} \ n_{max,V}\ge 1, \\ \qquad V_{reg} \qquad \qquad \qquad \qquad \qquad  \text{for} \ n_{max,V}=0,
\end{cases}   \label{Vpoleser}
\end{align}
where %
\begin{equation} \label{Rreg}
R_{reg}=\sum\limits_{j=0}^{\infty}{R_{j}}{(t)(w - a)^j}
\end{equation}
and
\begin{equation} \label{Vreg}
V_{reg}=\sum\limits_{j=0}^{\infty}{V_{j}}{(t)(w - a)^j}
\end{equation}
 are the regular parts of $R$ and $V$ (these regular parts are the analytic functions at
$w=a(t)$). It is assumed that $R_{-n_{max,R}}(t)$ and $V_{-n_{max,V}}(t)$ are nonzero.  We also define the Taylor series representations at $w=a$  of the functions $\bar R$ and $\bar V$ (these functions are analytic  at $w=a$  from the definition \e{bardef} because both  $R$\ and $V$ are analytic at $w=\bar a$)   as follows%
\begin{equation} \label{Rbarexpnasion}
\bar R(w,t)\equiv R_c(t)+\sum\limits_{j=1}^{\infty}{R_{c,j}}{(t)(w - a)^j}.
\end{equation}
and
\begin{equation} \label{Vbarexpnasion}
\bar V(w,t)\equiv V_c(t)+\sum\limits_{j=1}^{\infty}{V_{c,j}}{(t)(w - a)^j},
\end{equation}
where   $R_c(t)=   R_{c,0}(t)\equiv\bar R(w,t)|_{w=a}$ and
$V_c(t)=V_{c,0}(t)\equiv\bar V(w,t)|_{w=a}$ are zero order terms and
${R_{c,j}}(t)$, ${V_{c,j}}(t)$ are the coefficients of the higher order
terms of the respective power series. Then Eqs. \e{Reqn4}-\e{Veqn4} can
have persistent in time pole solution \e{Rpoleser}-\e{Vreg}, such that
both $R$ and $V$ have only simple poles singularities at a moving point
$w=a(t) $ only if the following conditions are all satisfied:

(a)  $n_{max,V}<n_{max,R}$, i.e. the order  of the highest  poles of $V$ is always lower than the  order $n_{max,R}$ of the highest pole of $R.$

(b) Moreover, $n_{max,V}\le n_{max,R}-m-1$, where $m=(n_{max,R}-2)/2$ for $n_{max,R}$ even and $m=(n_{max,R}-1)/2$ for $n_{max,R}$ odd.

(c) The coefficients of  equation \e{Vbarexpnasion} must satisfy the conditions  $V_{c,1}=V_{c,2}=\ldots=V_{c,m}=0$ provided $n_{max,R}\ge 3$,
 where $m$ is defined in (b).

(d) The coefficient of the  highest nonzero pole of $V$ is given by     $V_{-n_{max,R}+m+1} =- \frac{R_{-n_{max,R}}V_{c,m+1}}{R_c} $\ provided $n_{max,R}\ge 2$,
 where $m$ is defined in (b).

\vspace{0.6cm}

{\it Remark 1}. For the particular case of $n_{max,R}=0$,  Theorem 1
recovers Theorem 1 of Ref.
\cite{DyachenkoDyachenkoLushnikovZakharovJFM2019}.

{\it Remark 2}. $R_c$ in the denominator in (d) does not create any problem because   the conformal map \e{zwdef} and the definition \e{RVvar1} imply that $R(w)\ne0$ for    $w\in\C^-$ and, respectively, %
\begin{equation} \label{Rneq0}
\bar R_c=R(w)|_{w=a}\ne0 \ \text{for}   \  a\in\C^+.
\end{equation}This is a fact of essential importance for the proof of Theorem 1.

{\it Remark 3}. In addition to the expression in (d) in Theorem 1, it is
possible to provide the explicit expressions for the  coefficients
$V_{-n_{max,R}+m+2},\ldots, V_{-1}$ provided $n_{max,R}\ge 4$. These
coefficients are fully determined by the coefficients in equations
\e{Rpoleser}, \e{Rbarexpnasion} and \e{Vbarexpnasion} only (and depend
neither  on time derivatives of these coefficients   or $a_t$ and $g$).
 In particular,
\begin{equation}\label{VnmaxRm2}
V_{-n_{max,R}+m+2} =\frac{R_{-n_{max,R}} (R_{c,1} V_{c,m+1}-R_c V_{c,m+2})-R_{-n_{max,R}+1}R_c V_{c,m+1}}{(R_c)^2}.
\end{equation}
  However, the other explicit expressions for  $V_{-n_{max,R}+m+3},\ldots, V_{-1}$ turn increasingly bulky with the increase of $n_{max,R}$ so we do not provide them here.
Eq.
\e{VnmaxRm2} is derived as  the byproduct of the proof of Theorem  1 in Section \ref{sec:poles}.

{\it Remark 4}. Theorem 1 provides only the necessary conditions for the
existence of the persistent pole solutions. These necessary conditions are
quite restrictive and
 it appears likely that except very rear exceptions such  persistent pole solutions do not exist. The only known exception is the trivial case
\begin{equation} \label{Vgtrivial}
g=0, \quad \frac{\partial R}{\partial t}\equiv 0, \quad  \text{and} \quad
V\equiv 0,
\end{equation}
 i.e. a stationary solution of fluid at rest without gravity. In equations  \e{Reqn4}-\e{Veqn4}, the zero velocity $V\equiv 0$ implies that $U=B\equiv 0$. Then  equation \e{Reqn4} is satisfied by  $\frac{\partial R}{\partial t}\equiv 0$ while equation \e{Veqn4} reduces  to $g(R-1)\equiv 0$. Then either $R\equiv 1$, i.e. a flat free surface (which we do not consider as absolutely trivial) or $g=0$ as in equation \e{Vgtrivial}.   Any singularity of $R$ for
 $w\in \C^+ $ are allowed for the stationary  solution \e{Vgtrivial}. In the sense of the existence of such trivial solution,
 Theorem 1 cannot be improved at least for $g=0$ to fully exclude pole solutions in $R.$

Another way to strengthen Theorem 1 is to address the existence of  the
purely rational time-dependent solutions of equations \e{Reqn4}-\e{Veqn4}.
It would be generally  extremely attractive to find rational solutions
containing only pole-type singularities in $w$. There are examples of
different reductions/models of free surface hydrodynamics which allows
such rational solutions. They include a free surface  dynamics  for the
quantum Kelvin-Helmholtz instability between two components of superfluid
Helium \cite{LushnikovZubarevPRL2018,LushnikovZubarevJETP2019}; an
interface dynamic between ideal fluid and light highly viscous
 fluid \cite{LushnikovPhysLettA2004}, and
a  motion of  the dielectric fluid with a charged and ideally conducting
free surface in the vertical electric field
\cite{Zubarev_JETPLett_2000,Zubarev_JETP_2002,ZubarevJETP2008eng}.
 The general case of the ideal fluid with free surface considered in this paper however appears to resists heavily to the existence of such rational solutions. The following theorem is proven in Section \ref{sec:Rationalsolution}:

{\it\ Theorem} 2. Assume the following rational solution of equations
\e{Reqn4}-\e{Veqn4}: %
\begin{equation}\label{VR12}
\begin{split}
&R= \frac{R_{-2}(t)}{(w-a(t))^2}+  \frac{R_{-1}(t)}{(w-a(t))}+1,\\
& V= \frac{V_{-1}(t)}{(w-a(t))}. \\
\end{split}
\end{equation}
Then beyond the trivial solution \e{Vgtrivial},  all possible solutions of equations \e{Reqn4}-\e{Veqn4} have one or two zeros of $R(w,t)$   either for $w\in\R$  or for $w\in\C^-$. It implies the singularity of the conformal map \e{zwdef} through the definition \e{RVvar1} contradicting the assumption of the mapping of   $\C^-$ into the area occupied by fluid.  Thus no non-trivial
rational solution  \e{VR12}  exists.
In other words, the explicit family of nontrivial rational
  solutions obtained in the proof of this theorem is nonphysical because of the violation of the condition  $R(w)\ne 0.$

\vspace{0.6cm}

{\it Remark 5}. The rational solution \e{VR12}   however satisfies  Theorem 1 by allowing up to the
second order pole in $R$ and the fist order pole in $V.$ This is the example that  Theorem 1 provides only necessary conditions for the existence of the solutions with poles in $R$\ and $V.$

{\it Remark 6}. The  last term in the right-hand side (r.h.s.) of the
first equation of  \e{VR12} is chosen to satisfy $R\to 1$ as required from
Eq. \e{zlimit} at $w\to\infty, \ w\in \C^-.$ Also the second equation in
\e{VR12} satisfies the decaying BC \e{Dirichlet2}.

{\it Remark 7}. Theorem 1 is the local results
  because we use the Laurent series of solutions of free surface hydrodynamics  at any moving point
   $w=a(t)$, $Im(a)>0$. It means that we are not restricted to rational solutions because such local analysis does not exclude the existence of branch points
    for $w\ne a(t), \ w\in \C^+$.    In contrast, Theorem 2 is the global results because it fully excludes the existence of the rational solution  \e{VR12} valid for any $w\in\C.$

{\it Remark 8}. The exact rational solutions  of Eqs. \e{Reqn4}-\e{Veqn4} were obtained in Refs. \cite{ZakharovDyachenkoConferenceTalk2016,ZubarevKarabutJETPLett2018eng,DyachenkoDyachenkoLushnikovZakharovJFM2019,ZakharovTMF2019accepted} for the non-decaying BCs, i.e. for the infinite  energy of the fluid.

In contrast to the solution with pole singularities, we show in Section
\ref{sec:Persistencebranchcuts} that power law branch points are
persistent with equations  \e{Reqn4}-\e{Veqn4} which is consistent with
previous results of Refs.
\cite{MalcolmGrantJFM1973LimitingStokes,TanveerProcRoySoc1991,TanveerProcRoySoc1993,KuznetsovSpektorZakharovPhysLett1993,KuznetsovSpektorZakharovPRE1994,MooreProcRSocLond1979,MeironBakerOrszagJFM1982,BakerMeironOrszagJFM1982,KrasnyJFM1986,CaflischOrellanaSIAMJMA1989,CaflischOrellanaSiegelSIAMJAM1990,BakerShelleyJFM1990,ShelleyJFM1992,CaflishEtAlCPAM1993,BakerCaflischSiegelJFM1993,CowleyBakerTanveerJFM1999,BakerXieJFluidMech2011,Zubarev_Kuznetsov_JETP_2014,KarabutZhuravlevaJFM2014,ZakharovDyachenkoConferenceTalk2016,ZubarevKarabutJETPLett2018eng,DyachenkoDyachenkoLushnikovZakharovJFM2019,ZakharovTMF2019accepted}
obtained by various analytic and numerical techniques.

\section{Non-persistence of poles in $R$ and $V$ variables}
\label{sec:poles}

In this section we prove Theorem 1. \begin{proof} We start the proof by
recalling Remark 1  that $R(w)\ne0$ for    $w\in\C^-$, see Eq. \e{Rneq0}.
Here and below we often omit the second argument $t$ when we focus on
analytical properties in $w$.

All four functions  $R$, $V$,  $U$ and $B$ of Eqs. \e{Reqn4}-\e{Veqn4} must have singularities in the upper half-plane
$w\in\C^+$ while being analytic for   $w\in\C^-$. To understand that consider  the Laurent series
\e{Rpoleser} and \e{Vpoleser} and, similar, the Laurent series \begin{align}\label{Upolen}
&U =\sum\limits_{j=-max(n_{max,V},n_{max,R})}^{-1}{U_j}{(w - a)^j}+ U_{reg}, \ \  U_{reg}=\sum\limits_{j=0}^{\infty}{U_{j}}{(w - a)^j},  \\
&B = \sum\limits_{j=-n_{max,V}}^{-1}{B_j}{(w - a)^j} + B_{reg}, \quad \qquad \qquad \ \  B_{reg}=\sum\limits_{j=0}^{\infty}{B_{j}}{(w - a)^j}.  \label{Bpolen}
\end{align}
To understand validity of these equations, we notice that using Eq. \e{Projectordef} we can rewrite the definitions \e{UBdef4} as
\begin{equation}\label{UBPplus}
\begin{split}
& U=R\bar V+\bar RV-\hat P^+(R\bar V+\bar RV), \\
& B=V\bar V-\hat P^+(V\bar V).
\end{split}
\end{equation}
The functions $\hat P^+(R\bar V+\bar RV)$ and  $\hat P^+(V\bar V)$ are analytic at $w=a\in\C^+$ thus they only contribute to the regular parts $U_{reg}$ and $B_{reg},$ respectively. The functions $\bar R$ and $\bar V$ are also analytic at $w=a$ with the Taylor series representations \e{Rbarexpnasion} and  \e{Vbarexpnasion}.
The sum of two terms $R\bar V+\bar RV$ in r.h.s. of the first Eq. in \e{UBPplus} also explains why the summation in r.h.s. of Eq. \e{Upolen} starts from the most singular term with $j=-max(n_{max,V},n_{max,R}).$
 Eqs. \e{Rpoleser},\e{Vpoleser},\e{Rbarexpnasion},\e{Vbarexpnasion} and \e{Upolen}-\e{UBPplus} imply that generally $U$ and $B$ have the same types of singularities as $R$ and $V$ except special cases when poles of  either $R$ or $V$  are canceled out.

If $n_{max,R}\le n_{max,V},$ then
  the most singular term in Eqs. \e{Reqn4}-\e{Veqn4}  is $- \I \,n_{max,V}R_cV^2_{- n_{max,V}}(w-a)^{-2 n_{max,V}-1}$ in r.h.s of Eq. \e{Veqn4}, where we
  used Eqs.   \e{Rpoleser}-\e{Vbarexpnasion} and \e{Upolen}-\e{UBPplus}. It implies that $V_{- n_{max,V}}=0$ and, respectively, we must set that  $n_{max,R}> n_{max,V}$ which completes the proof of the statement (a) of Theorem 1 as well as it fully covers Theorem 1 for $n_{max,R}=1$ so in the remaining part of the proof we assume $n_{max,R}\ge 2.$
Also the power of the most singular term in Eq. \e{Upolen} turns into $j=-max(n_{max,V},n_{max,R})=-n_{max,R}.$

The most singular terms in the left-hand side  (l.h.s.) of  equations
\e{Reqn4} and \e{Veqn4}
result from the differentiation of $a$\ over $t$ and they have the orders $(w-a)^{-n_{max,R}-1}$  and $(w-a)^{-n_{max,V}-1}$, respectively. Thus they can be ignored for the leading orders analysis because they are much less singular than the leading terms in r.h.s. of these equations.

   The term of the order $(w-a)^{-2 n_{max,R}}$ is identically zero in Eq. \e{Veqn4}     because of   $n_{max,R}> n_{max,V.}$ Now the most singular term is  $- \I R_{-n_{max,R}}[ R_cV_{-n_{max,R}+1}+ R_{-n_{max,R}}V_{c,1}] \,(w-a)^{-2 n_{max,R}}$   in r.h.s of Eq. \e{Reqn4} which results in
\begin{equation}\label{nmaxR2}
V_{-n_{max,R}+1}= -R_{-n_{max,R}}V_{c,1}/R_c,  \ n_{max,R}\ge 2,
\end{equation}
 because $R_{-n_{max,R}}\ne0$ by the assumptions of Theorem 1. For $n_{max,R}=2$ Eq. \e{nmaxR2} completes the proof of Theorem 1 so  in the remaining part of the proof we assume $n_{max,R}\ge 3.$

Using Eq. \e{nmaxR2} to exclude  $V_{-n_{max,R}+1}$, we obtain the next
order term in r.h.s of Eq. \e{Veqn4}  as   $\frac{- \I(n_{max_R}-2)
R_{-n_{max,R}}^2V_{c,1}^2}{R_c} \,(w-a)^{-2 n_{max,R}+1}$ which
      implies that%
\begin{equation}\label{Vc1n3}
V_{c,1}=0,  \ n_{max,R}\ge 3.
\end{equation}
  But then Eq. \e{nmaxR2} results in%
\begin{equation}\label{VnR1n3}
V_{-n_{max,R}+1}=0,  \ n_{max,R}\ge 3.
\end{equation}
  Thus we must set%
\begin{equation}\label{nVnRm2}
n_{max,V}\le n_{max,R}-2,  \ n_{max,R}\ge 3
\end{equation}
 which recovers the statement (b) of Theorem 1 for both $n_{max,R}=3 $ and  $n_{max,R}=4 $ ($m=1$ in both these cases as follows from the definition of $m$ in the statement of Theorem 1).   From Eqs. \e{Vc1n3} and \e{VnR1n3} we obtain that the most singular term in r.h.s of Eq. \e{Reqn4}  is

  $-2 \I R_{-n_{max,R}}[ R_cV_{-n_{max,R}+2}+ R_{-n_{max,R}}V_{c,2}] \,(w-a)^{-2 n_{max,R}+1}$     which results in
\begin{equation}\label{nmaxR3}
V_{-n_{max,R}+2}= -R_{-n_{max,R}}V_{c,2}/R_c,  \ n_{max,R}\ge 3,
\end{equation}
 because $R_{-n_{max,R}}\ne0$ by the assumptions of Theorem 1. For both $n_{max,R}=3$ an  $n_{max,R}=4$, Eqs. \e{Vc1n3}-\e{nmaxR3} complete the proof of Theorem 1.
So  in the remaining part of the proof we assume that $n_{max,R}\ge 5.$

{\it\ Remark 9.} For  $n_{max,R}\ge 4$ one can consider  at least one next order before reaching terms with $A_t$ in l.h.s..
Then the term of the order $(w-a)^{-2 n_{max,R}+2}$ is identically zero in
Eq. \e{Veqn4}     because of Eqs.  \e{nmaxR2}-\e{nmaxR3}. Now the most
singular term is  $\propto (w-a)^{-2 n_{max,R}+2}$   in r.h.s of Eq.
\e{Reqn4} which results in Eq. \e{VnmaxRm2} from Remark 3 for $
n_{max,R}=4$ and, respectively, $m=2.$

Proceeding further by induction for  $n_{max,R}\ge 5,$ we complete the proof
 of  Theorem 2 through straightforward calculations by collecting the remaining  terms of powers $(w-a)^{-2n_{max,R}+2}, \ldots      ,(w-a)^{-n_{max,R}-2} $ in Eqs. \e{Reqn4} and \e{Veqn4}. As it is seen from the previous steps of the induction, the even and odd values of   $n_{max,R}$ need to be treated a little differently. For the odd values one has to take into account all terms of powers  $(w-a)^{-2n_{max,R}+2}, \ldots      ,(w-a)^{-n_{max,R}-2}. $ For the even values it is sufficient to take into account only terms of powers  $(w-a)^{-2n_{max,R}+2}, \ldots      ,(w-a)^{-n_{max,R}-3} .$ The extra term of the power $(w-a)^{-n_{max,R}-2}$ is identically zero in Eq. \e{Veqn4} while the term of the same power in Eq. \e{Reqn4} can be used to find the expression for $V_{-1}$. In contrast, for the odd values of   $n_{max,R},$  it is necessary to take into account all terms of powers  $(w-a)^{-2n_{max,R}+2}, \ldots      ,(w-a)^{-n_{max,R}-2} .$ Then the  term of the power $(w-a)^{-n_{max,R}-2}$  in Eq. \e{Veqn4} ensures that $V_{c,m}=0$ as required in the statement (c) of Theorem 1 while the term of the same power in Eq. \e{Reqn4} can be used to find the expression for $V_{-1}$. This concludes the proof of Theorem 1.

~~~~~~~~~~~~~~~~~~~~~~~~~~~~~~~~~~~~~~~~~~~~~~~~~~~~~~~~~~~~~~~~~~~~~~~~$\qquad\qquad\square$
\end{proof}

{\it\ Remark 10.} One can immediately count that the total number of
conditions obtained   from the powers    $(w-a)^{-2n_{max,R}-1}, \ldots
,(w-a)^{-n_{max,R}-2} $ in  Eqs. \e{Reqn4} and \e{Veqn4} is  $2n_{max,R}.
$ However, the  number of the nontrivial conditions is only $n_{max,R}+m.$
These nontrivial conditions results in
$V_{-n_{max,R}}=V_{-n_{max,R}+1}=\ldots=V_{-n_{max,R}+m}=0$,
$V_{c,1}=\ldots=V_{c,m}=0$, and explicit expressions for
$V_{-n_{max,R}+m+1},\ldots, V_{-1}$.    Remaining trivially satisfied
$2n_{max,R}-m$ conditions (trivial zeros) occur in  Eq. \e{Veqn4} for the
terms with the powers
$(w-a)^{-2n_{max,R}},(w-a)^{-2n_{max,R}+2},(w-a)^{-2n_{max,R}+4}, \ldots
,(w-a)^{-n_{max,R}-2} $ for the even  $n_{max,R}$  and for the  powers
$(w-a)^{-2n_{max,R}},(w-a)^{-2n_{max,R}+2},(w-a)^{-2n_{max,R}+4}, \ldots
,(w-a)^{-n_{max,R}-3} $ for the odd  $n_{max,R}.$ Another trivial zero is
for  the power     $(w-a)^{-2n_{max,R}-1} $    in  Eq. \e{Reqn4}.  Terms
of lower orders $(w-a)^{-n_{max,R}-1}, \ldots      $ can be additionally
used to provide conditions for time derivative of different coefficients.
Roughly we can summarize Theorem 1 that the order of poles in $V$ is at
least twice smaller than the order of poles in $R$

 %

%In conclusion of this section, we note that taking into account a nonzero %surface tension, i.e. working with Eqs.  \e{Udef2},\e{Bintdef2}, \e{Reqn} %and \e{Veqn}
% instead of Eqs. \e{Reqn4}-\e{Veqn4}, immediately shows that pole singularity %both for $R$ and $V$ is non-persistent because the dependence of surface %tension terms of $Q=\sqrt{R}$  introduces the square root singularity into %Eq.  \e{Veqn}
%  which cannot be compensated by other terms with poles.

\section{Nonexistence of the rational solution with the first or the second order poles}
\label{sec:Rationalsolution}

In this section we prove Theorem 2.  We  first obtain the exact rational solution of  Eqs. \e{Reqn4}-\e{Veqn4} but then show that it is not physical. %\subsection{Example of purely rational %solution}

\begin{proof} We look for all possible functions $R_{-1}(t)$, $R_{-2}(t)$, $V_{-1}(t)$ and $a(t)$  such that Eq.  \e{VR12}  is the exact solution of Eqs. \e{Reqn4}-\e{Veqn4}.
We plug in    \e{VR12}  into Eqs.
\e{Reqn4}-\e{Veqn4} and look for the exact solutions. The projectors
in Eq. \e{UBdef4} are easy to evaluate using partial fractions over
$w$ if we notice that the complex conjugation of Eq. \e{VR12}  is
given by
\begin{equation}\label{VR12conj}
\begin{split}
&\bar R= \frac{\bar R_{-2}(t)}{(w-\bar a(t))^2}+  \frac{\bar R_{-1}(t)}{(w-\bar a(t))}+1,\\
& \bar V= \frac{\bar V_{-1}(t)}{(w-\bar a(t))}, \\
\end{split}
\end{equation} where we recall that we do not conjugate $w$ to obtain the analytical continuation from the real line $w=u$ as explained in Section \ref{sec:Introduction}.

We collect terms with all possible powers of $(w-a)$  in both Eqs.
\e{Reqn4} and \e{Veqn4}. The order $(w-a)^{-5}$ is trivially satisfied
because we set  $V_{-2}=0$ in Eq.  \e{VR12}
as required by Theorem 1. The order $(w-a)^{-4}$ needs that%
\begin{equation} \label{V1R2order4}
V_{-1}=\frac{R_{-2} \bar V_{-1}}{\bar R_{-2}+(a-\bar a) (\bar
R_{-1}+a-\bar a)},
\end{equation}
where we assumed that $V_{-1}\ne 0.$ In the opposite case of
$V_{-1}= 0,$ we immediately obtain that the only possible solution
is $g=0$ and both $R_{-2},R_{-1}$ are time independent thus
recovering the trivial case \e{Vgtrivial}.  Thus below we assume
$V_{-1}\ne 0$.

The order $(w-a)^{-3}$ in Eq. \e{Veqn4} is satisfied by Eq. \e{V1R2order4} while Eq. \e{Reqn4} requires that%
\begin{equation} \label{V1R2order3}
a_t=\frac{\I \bar V_{-1}}{a-\bar a}.
\end{equation}

The order $(w-a)^{-2}$ in Eq. \e{Reqn4}  together with the  condition \e{V1R2order4}  requires a time independence of $R_{-2}$, i.e. %
\begin{equation} \label{V1R2order2}
R_{-2}=const
\end{equation}%
 (provided $R_{-2}\ne 0$)  while Eq. \e{Veqn4} at that order is valid only for
\begin{equation} \label{g0eq}
g=0.
\end{equation}
The order $(w-a)^{-1}$ in Eq.  \e{Reqn4} requires a time independence of $R_{-1}$, i.e. %
\begin{equation} \label{V1R2order1}
R_{-1}=const
\end{equation}%
  while  Eq. \e{Veqn4} needs a time independence of $V_{-1}$, i.e. %
\begin{equation} \label{V1R2order1a}
V_{-1}=const.
\end{equation}

Solving Eq.  \e{V1R2order4}  for $(a-\bar a)$ together with Eqs.
\e{V1R2order2}, \e{V1R2order1}
and \e{V1R2order1a} show that $(a-\bar a)$ must be constant in time, i.e. the imaginary part of $a$ must be constant. Then Eq. \e{V1R2order3} requires that $V_{-1}=Re(V_{-1})$ and, moreover,   %
\begin{equation} \label{RVparam1}
a=a_{r,1}t+a_{r,0}+\I a_i, \ V_{-1}=2 a_{r,1}a_i, \
\end{equation}where  $a_{r,1}, a_{r,0}, a_i$ are the arbitrary  real constants. It remains to satisfy Eq.  \e{V1R2order4} which together with Eq. \e{RVparam1} gives that
either $a_{r,1}=V_{-1}=0$ (which recovers the trivial case
\e{Vgtrivial}) or   \begin{equation} \label{RVparam1a}
 R_{-1}=\frac{Im(R_{-2})}{a_i}+2\I a_i.
\end{equation}
The exact solution  \e{VR12}, \e{RVparam1}, \e{RVparam1a} is valid
for the arbitrary complex constant value of $R_{-2}$ and zero gravity $g=0.$ It means that solution
propagates with the constant velocity in the horizontal direction
with all residues being time independent. % which recovers the% solution
%\e{VR12} and \e{RVparam}.

The analyticity of $R$ for $w\in\C^-$
requires that $a_i>0.$
We now check locations of zeros of $R$ which are poles of $z_u.$ Using Eq. \e{VR12} we obtain that  $R=0$ for %
\begin{equation} \label{weq0}
w=a_{r,1}t+a_{r,0}-\frac{Im(R_{-2})}{2a_i}\pm\left (
\frac{Im(R_{-2})^2}{4a_i^2}-a_i^2-Re(R_{-2})\right )^{1/2}.
\end{equation}
Eq. \e{weq0} either has two real roots which implies a singularity
at fluid's free surface with mapping of $z(w)$ into infinity or it
has two complex conjugated roots, one is in $\C^-$ thus violating the
analyticity of $z(w)$ for $w\in\C^-$. Thus we conclude that the
rational solution  \e{VR12} is not compatible with
the condition \e{zwconformal} that the  mapping \e{zwdef} is conformal for  $w\in\C^-$ which completes the proof.

\end{proof}

~~~~~~~~~~~~~~~~~~~~~~~~~~~~~~~~~~~~~~~~~~~~~~~~~~~~~~~~~~~~~~~~~~~~~~~~~~~~~~$\square $

\section{Persistence of branch cuts}
\label{sec:Persistencebranchcuts}

We show in this Section  that, contrary to poles analyzed in Section \ref{sec:poles}, power law branch cuts are persistent in time for free surface dynamics.   Assume that in the small neighborhood of $w = a$, the following
expansions hold
\begin{equation}\label{VRalpha}
\begin{split}
V &= V_0 + V_\gamma(w-a)^\gamma + \ldots, \\
R &= R_0 + R_\gamma(w-a)^\gamma + \ldots, \\
U &= U_0 + U_\gamma(w-a)^\gamma + \ldots, \\
B &= B_0 + B_\gamma(w-a)^\gamma + \ldots,
\end{split}
\end{equation}
where $\gamma$ is the complex number and $``\ldots"$ designates terms
with less singular powers (i.e. with powers $\gamma_1$ such that $Re(\gamma)<Re(\gamma_1)).$ Similar to Section \ref{sec:poles}, we perform local
analysis  at $w=a$ on the persistence of singularities but this time with the expansion \e{VRalpha}.

Eqs. \e{UBPplus} and  \e{VRalpha}  imply that \begin{align} U_\gamma
= R_c V_\gamma + V_c  R_\gamma, \,\, \mbox{and} \,\, B_\gamma = V_c
V_\gamma, \label{eqn13}
\end{align}
where we collected terms with the power  $(w-a)^\gamma$ and used
definitions of $R_c$ and $V_c$ from Eqs. \e{Rbarexpnasion} and  \e{Vbarexpnasion}.

Plugging expansions \e{VRalpha} into  Eqs. \e{Reqn4}-\e{Veqn4} above
and collecting the most singular terms of the order
$(w-a)^{\gamma-1}$, we obtain that
\begin{align}\label{atcond1}
-R_\gamma\frac{\partial a }{\partial t} &= \I \left(U_0 R_\gamma- R_0 U_\gamma  \right), \\
-V_\gamma\frac{\partial a }{\partial t} &= \I \left(U_0 V_\gamma-
R_0 B_\gamma  \right). \label{atcond2}
\end{align}
Multiplying Eq. \e{atcond1} by $V_\gamma$ and subtracting from it
Eq. \e{atcond2} multiplied by  $R_\gamma$, we obtain the
compatibility condition
\begin{align}\label{UalphaB1}
R_0(U_\gamma V_\gamma- B_\gamma R_\gamma) = 0.
\end{align}
Using Eqs.  \e{UalphaB1} and ~\eqref{eqn13} we find the compatibility condition%
\begin{equation} \label{R0RcVc}
R_0 R_c V_\gamma  = 0.
\end{equation}
According to our assumptions  $R_c \neq 0 $ as explained in Section
\ref{sec:poles}.  Then the remaining possibilities in Eq. \e{R0RcVc}
are that either $R_0=0$ or $V_\gamma=0$.  The first possibility is
that we assume that $V_\gamma\ne 0$ which implies that\begin{align}
R_0 = 0. \label{eqn16}
\end{align}
Then Eqs. \e{atcond1} and \e{atcond2} result in a simple equation for the singularity location \begin{align}\label{atcond3}
\frac{\partial a }{\partial t} &=- \I U_0.
\end{align}
Eq. \e{eqn16} means that branch points are zeros of the function
$R$.  There is no restriction on the value of $\gamma$ which  is a
predicament to persistence of branch points of arbitrary types.
Nevertheless the most common type of branch points, observed in  our
numerical experiments is $\gamma = \frac{1}{2}$ which is consistent
with the results of Refs.
\cite{MalcolmGrantJFM1973LimitingStokes,TanveerProcRoySoc1991,TanveerProcRoySoc1993,KuznetsovSpektorZakharovPhysLett1993,KuznetsovSpektorZakharovPRE1994}.
Square root singularities have been also intensively studied based
on the representation of vortex sheet in Ref.
\cite{MooreProcRSocLond1979,MeironBakerOrszagJFM1982,BakerMeironOrszagJFM1982,KrasnyJFM1986,CaflischOrellanaSIAMJMA1989,CaflischOrellanaSiegelSIAMJAM1990,BakerShelleyJFM1990,ShelleyJFM1992,CaflishEtAlCPAM1993,BakerCaflischSiegelJFM1993,CowleyBakerTanveerJFM1999,BakerXieJFluidMech2011,Zubarev_Kuznetsov_JETP_2014,KarabutZhuravlevaJFM2014,ZubarevKarabutJETPLett2018eng}.

Particular solution of  Eqs. \e{Reqn4}-\e{Veqn4}  is Stokes wave which is
a nonlinear periodic gravity wave propagating with the constant velocity
~\cite{Stokes1847,Stokes1880}. In the generic situation, when the
singularity of Stokes wave is away from the real axis (non-limiting Stokes
wave), the only allowed singularity in $\C$ is $\gamma=1/2$ as was proven
in Ref. \cite{TanveerProcRoySoc1991} for the first (physical) sheet  of
the Riemann surface and  in Ref. \cite{LushnikovStokesParIIJFM2016} for
the infinite number of other (non-physical) sheets of Riemann surface.
Refs.
\cite{DyachenkoLushnikovKorotkevichJETPLett2014,DyachenkoLushnikovKorotkevichPartIStudApplMath2016,LushnikovDyachenkoSilantyevProcRoySocA2017}
provided detailed numerical verification of these singularities. The
limiting Stokes wave is the special case   $\gamma=1/3$  with $a=\I
Im(a)$. Also Ref.  \cite{TanveerProcRoySoc1993} suggested the possibility
in exceptional cases of the existence of $\gamma=1/n$ singularities with
$n$ being any positive integer as well as singularities involving
logarithms.

  The second possibility to satisfy the compatibility
  condition \e{R0RcVc} is to assume that $V_\gamma=0$. In that case either $V$ is the regular function at $w=a$ (while $R$ has a
  branch point at $w=a$) or  one of less singular terms is not zero. We also notice that $(R_0)_t\propto R_0$ (i.e. the case \e{eqn16} corresponds to the zero initial condition for $R_0$) as can be obtained from the analysis similar to provided above in this section. A further study of that case is beyond of the scope of this paper.
  %In the  section \e{sec:narrowbranchcut}??? we provide an example of the case of a regular $V.$

 %We prove in Appendix \ref{sec:AppA} %that for $g\ne 0 $  such type of solutions does not exist while for $g=0$ %the only possible family of solutions (beyond  the trivial case \e{Vgtrivial}) %is given by   %
%\begin{equation} \label{RVparam}
%a=a_{r,1}t+a_{r,0}+\I a_i, \ V_{-1}=2 a_{r,1}a_i, \
%R_{-1}=\frac{Im(R_{-2})}{a_i}+2\I a_i,
%\end{equation}
%where  $a_{r,1}, a_{r,0}, a_i$ are the arbitrary  real constants and
%$R_{-2}$ is the arbitrary complex constant.
%It means that solution
%propagates with the constant velocity in the horizontal direction
%with all residues being time independent. However, we show in
%Appendix \ref{sec:AppA} that  the rational solution  \e{VR12} and
%\e{RVparam} is not compatible with the condition that the  mapping
%\e{zwdef} is conformal for $w\in\C^-$ because for any values
%$a_{r,1}, a_{r,0}, a_i,$  $R_{-2}$  it always  has singularity
%either on the real line $w=0$ or for $w\in\C^-.$

\section*{Conflict of interest}
 The authors declare that they have no conflict of interest.

\begin{acknowledgements}

 The  work of P.M.L.   and V.E.Z.  was   supported by the state assignment ``Dynamics of the complex systems".
 The  work of
P.M.L.  was   supported by the National Science Foundation, grant
DMS-1814619. The work of V.E.Z. was supported by the National Science
Foundation, grant number DMS-1715323.
 The  work of V.E.Z. described in section 2 was supported by the Russian Science Foundation, grant
number 19-72-30028.

\end{acknowledgements}


\bibliographystyle{spmpsci}      % mathematics and physical sciences

%\bibliography{surfacewaves,lushnikov,biblionls,books,helium}

\begin{thebibliography}{10}
\providecommand{\url}[1]{{#1}} \providecommand{\urlprefix}{URL }
\expandafter\ifx\csname urlstyle\endcsname\relax
  \providecommand{\doi}[1]{DOI~\discretionary{}{}{}#1}\else
  \providecommand{\doi}{DOI~\discretionary{}{}{}\begingroup
  \urlstyle{rm}\Url}\fi

\bibitem{BakerCaflischSiegelJFM1993}
Baker, G., Caflisch, R.E., Siegel, M.: {Singularity formation during
  Rayleigh--Taylor instability}.
\newblock {Journal of Fluid Mechanics} \textbf{252}, 51--78 (1993)

\bibitem{BakerMeironOrszagJFM1982}
Baker, G.R., Meiron, D.I., Orszag, S.A.: {Generalized vortex methods for
  free-surface flow problems}.
\newblock {Journal of Fluid Mechanics} \textbf{123}, 477--501 (1982)

\bibitem{BakerShelleyJFM1990}
Baker, G.R., Shelley, M.J.: {On the connection between thin vortex layers
and
  vortex sheets}.
\newblock {Journal of Fluid Mechanics} \textbf{215}, 161--194 (1990)

\bibitem{BakerXieJFluidMech2011}
Baker, G.R., Xie, C.: Singularities in the complex physical plane for deep
  water waves.
\newblock J. Fluid Mech. \textbf{685}, 83--116 (2011)

\bibitem{CaflischOrellanaSIAMJMA1989}
Caflisch, R., Orellana, O.: {Singular Solutions and Ill--Posedness for the
  Evolution of Vortex Sheets}.
\newblock SIAM Journal on Mathematical Analysis \textbf{20}(2), 293--307 (1989)

\bibitem{CaflischOrellanaSiegelSIAMJAM1990}
Caflisch, R., Orellana, O., Siegel, M.: {A Localized Approximation Method
for
  Vortical Flows}.
\newblock SIAM Journal on Applied Mathematics \textbf{50}(6), 1517--1532 (1990)

\bibitem{CaflishEtAlCPAM1993}
Caflisch, R.E., Ercolani, N., Hou, T.Y., Landis, Y.: {Multi-valued
solutions
  and branch point singularities for nonlinear hyperbolic or elliptic systems}.
\newblock Communications on Pure and Applied Mathematics \textbf{46}(4),
  453--499 (1993)

\bibitem{ChalikovSheininAdvFluidMech1998}
Chalikov, D., Sheinin, D.: {Direct modeling of one-dimensional nonlinear
  potential waves}.
\newblock {Adv. Fluid Mech} \textbf{17}, 207--258 (1998)

\bibitem{ChalikovSheininJCompPhys2005}
Chalikov, D., Sheinin, D.: {Modeling of extreme waves based on equation of
  potential flow with a free surface}.
\newblock {Journal of Computational Physics} \textbf{210}, 247--273 (2005)

\bibitem{ChalikovBook2016}
Chalikov, D.V.: Numerical Modeling of Sea Waves.
\newblock Springer (2016)

\bibitem{CowleyBakerTanveerJFM1999}
Cowley, S.J., Baker, G.R., Tanveer, S.: {On the formation of Moore
curvature
  singularities in vortex sheets}.
\newblock J. Fluid Mech. \textbf{378}, 233--267 (1999)

\bibitem{CraigSulemJCompPhys1993}
Craig, W., Sulem, C.: Numerical simulation of gravity waves.
\newblock J. Comput. Phys. \textbf{108}, 73--83 (1993)

\bibitem{Dyachenko2001}
Dyachenko, A.I.: On the dynamics of an ideal fluid with a free surface.
\newblock Dokl. Math. \textbf{63}(1), 115--117 (2001)

\bibitem{DyachenkoDyachenkoLushnikovZakharovJFM2019}
Dyachenko, A.I., Dyachenko, S.A., Lushnikov, P.M., Zakharov, V.E.:
{Dynamics of
  Poles in 2D Hydrodynamics with Free Surface: New Constants of Motion}.
\newblock {Journal of Fluid Mechanics} \textbf{874}, 891--925 (2019)

\bibitem{DKSZ1996}
Dyachenko, A.I., Kuznetsov, E.A., Spector, M., Zakharov, V.E.: Analytical
  description of the free surface dynamics of an ideal fluid (canonical
  formalism and conformal mapping).
\newblock Phys.\,Lett.\,A \textbf{221}, 73--79 (1996)

\bibitem{DyachenkoLushnikovZakharovJFM2019}
Dyachenko, A.I., Lushnikov, P.M., Zakharov, V.E.: {Non-canonical
Hamiltonian
  structure and Poisson bracket for two-dimensional hydrodynamics with free
  surface}.
\newblock {Journal of Fluid Mechanics} \textbf{869}, 526--552 (2019)

\bibitem{DyachenkoLushnikovKorotkevichJETPLett2014}
Dyachenko, S.A., Lushnikov, P.M., Korotkevich, A.O.: {The complex
singularity
  of a Stokes wave}.
\newblock JETP Letters \textbf{98}(11), 675--679 (2013).
\newblock \doi{10.7868/S0370274X13230070}

\bibitem{DyachenkoLushnikovKorotkevichPartIStudApplMath2016}
Dyachenko, S.A., Lushnikov, P.M., Korotkevich, A.O.: {Branch Cuts of
Stokes
  Wave on Deep Water. Part I: Numerical Solution and Pad\'e Approximation}.
\newblock {Studies in Applied Mathematics} \textbf{137}, 419--472 (2016).
\newblock \doi{DOI: 10.1111/sapm.12128}

\bibitem{MalcolmGrantJFM1973LimitingStokes}
Grant, M.A.: The singularity at the crest of a finite amplitude
progressive
  {Stokes} wave.
\newblock J. Fluid Mech. \textbf{59(2)}, 257--262 (1973)

\bibitem{KarabutZhuravlevaJFM2014}
Karabut, E.A., Zhuravleva, E.N.: {Unsteady flows with a zero acceleration
on
  the free boundary}.
\newblock J. Fluid Mech. \textbf{754}, 308--331 (2014)

\bibitem{KrasnyJFM1986}
Krasny, R.: {A study of singularity formation in a vortex sheet by the
  point--vortex approximation}.
\newblock Journal of Fluid Mechanics \textbf{167}, 65--93 (1986)

\bibitem{KuznetsovSpektorZakharovPhysLett1993}
Kuznetsov, E., Spector, M., Zakharov, V.: Surface singularities of ideal
fluid.
\newblock Physics Letters A \textbf{182}(4-6), 387 -- 393 (1993).
\newblock \doi{http://dx.doi.org/10.1016/0375-9601(93)90413-T}

\bibitem{KuznetsovSpektorZakharovPRE1994}
Kuznetsov, E.A., Spector, M.D., Zakharov, V.E.: Formation of singularities
on
  the free surface of an ideal fluid.
\newblock Phys. Rev. E \textbf{49}, 1283--1290 (1994).
\newblock \doi{10.1103/PhysRevE.49.1283}.
\newblock \urlprefix\url{http://link.aps.org/doi/10.1103/PhysRevE.49.1283}

\bibitem{LandauLifshitzHydrodynamics1989}
Landau, L.D., Lifshitz, E.M.: Fluid Mechanics, Third Edition: Volume 6.
\newblock Pergamon, New York (1989)

\bibitem{LushnikovPhysLettA2004}
Lushnikov, P.M.: Exactly integrable dynamics of interface between ideal
fluid
  and light viscous fluid.
\newblock Physics Letters A \textbf{329}, 49 -- 54 (2004)

\bibitem{LushnikovStokesParIIJFM2016}
Lushnikov, P.M.: {Structure and location of branch point singularities for
  Stokes waves on deep water}.
\newblock {Journal of Fluid Mechanics} \textbf{800}, 557--594 (2016)

\bibitem{LushnikovDyachenkoSilantyevProcRoySocA2017}
Lushnikov, P.M., Dyachenko, S.A., Silantyev, D.A.: {New conformal mapping
for
  adaptive resolving of the complex singularities of Stokes wave}.
\newblock {Proc. Roy. Soc. A} \textbf{473}, 20170198 (2017)

\bibitem{LushnikovZubarevPRL2018}
Lushnikov, P.M., Zubarev, N.M.: {Exact solutions for nonlinear development
of a
  Kelvin-Helmholtz instability for the counterflow of superfluid and normal
  components of Helium II}.
\newblock {Phys. Rev. Lett.} \textbf{120}, 204504 (2018)

\bibitem{LushnikovZubarevJETP2019}
Lushnikov, P.M., Zubarev, N.M.: {Explosive Development of the
Kelvin-Helmholtz
  Quantum Instability on the He-II Free Surface}.
\newblock {J. Exp. Theor. Phys.} \textbf{129}, 651--658 (2019)

\bibitem{MeironBakerOrszagJFM1982}
Meiron, D.I., Baker, G.R., Orszag, S.A.: {Analytic structure of vortex
sheet
  dynamics. Part 1. Kelvin--Helmholtz instability}.
\newblock Journal of Fluid Mechanics \textbf{114}, 283--298 (1982)

\bibitem{MeisonOrzagIzraelyJCompPhys1981}
Meison, D., Orzag, S., Izraely, M.: Applications of numerical conformal
  mapping.
\newblock J. Comput. Phys. \textbf{40}, 345--360 (1981)

\bibitem{MooreProcRSocLond1979}
Moore, D.W.: The spontaneous appearance of a singularity in the shape of
an
  evolving vortex sheet.
\newblock Proceedings of the Royal Society of London A: Mathematical, Physical
  and Engineering Sciences \textbf{365}(1720), 105--119 (1979).
\newblock \doi{10.1098/rspa.1979.0009}

\bibitem{Ovsyannikov1973}
Ovsyannikov, L.V.: Dynamics of a fluid.
\newblock M.A. Lavrent'ev Institute of Hydrodynamics Sib. Branch USSR Ac. Sci.
  \textbf{15}, 104--125 (1973)

\bibitem{ShelleyJFM1992}
Shelley, M.J.: {A study of singularity formation in vortex--sheet motion
by a
  spectrally accurate vortex method}.
\newblock Journal of Fluid Mechanics \textbf{244}, 493--526 (1992).
\newblock \doi{10.1017/S0022112092003161}

\bibitem{Stokes1847}
Stokes, G.G.: On the theory of oscillatory waves.
\newblock Transactions of the Cambridge Philosophical Society \textbf{8},
  441--455 (1847)

\bibitem{Stokes1880}
Stokes, G.G.: On the theory of oscillatory waves.
\newblock Mathematical and Physical Papers \textbf{1}, 197--229 (1880)

\bibitem{TanveerProcRoySoc1991}
Tanveer, S.: {Singularities in water waves and Rayleigh-Taylor
instability}.
\newblock Proc. R. Soc. Lond. A \textbf{435}, 137--158 (1991)

\bibitem{TanveerProcRoySoc1993}
Tanveer, S.: {Singularities in the classical Rayleigh-Taylor flow:
formation
  and subsequent motion}.
\newblock Proc. R. Soc. Lond. A \textbf{441}, 501--525 (1993)

\bibitem{WeinsteinJDiffGeom1983}
Weinstein, A.: {The local structure of Poisson manifolds}.
\newblock J. Differential Geometry \textbf{18}, 523--557 (1983)

\bibitem{ZakharovTMF2019accepted}
Zakharov, V.E.: {Integration of equations of deep fluids with free
surface}.
\newblock {Accepted to Theoretical and Mathematical Physics}  (2019)

\bibitem{ZakharovDyachenkoArxiv2012}
Zakharov, V.E., Dyachenko, A.I.: Free-surface hydrodynamics in the
conformal
  variables.
\newblock arXiv:1206.2046  (2012)

\bibitem{ZakharovDyachenkoConferenceTalk2016}
Zakharov, V.E., Dyachenko, A.I.: {Are equations of a deep fluid with a
free
  surface integrable?}
\newblock {Presentation at The Russian-French Symposium ``Mathematical
  Hydrodynamics", Abstracts, LIH SB RAS \& NSU, Novosibirsk, Russia, 55}
  (2016)

\bibitem{ZakharovDyachenkoVasilievEuropJMechB2002}
Zakharov, V.E., Dyachenko, A.I., Vasiliev, O.A.: New method for numerical
  simulation of nonstationary potential flow of incompressible fluid with a
  free surface.
\newblock European Journal of Mechanics B/Fluids \textbf{21}, 283--291 (2002)

\bibitem{Zubarev_JETPLett_2000}
Zubarev, N.M.: Charged-surface instability development in liquid helium:
An
  exact solution.
\newblock JETP Lett. \textbf{71}, 367--369 (2000)

\bibitem{Zubarev_JETP_2002}
Zubarev, N.M.: Exact solutions of the equations of motion of liquid helium
with
  a charged free surface.
\newblock J. Exp. Theor. Phys. \textbf{94}, 534--544 (2002)

\bibitem{ZubarevJETP2008eng}
Zubarev, N.M.: {Formation of Singularities on the Charged Surface of a
  Liquid-Helium Layer with a Finite Depth}.
\newblock Journal of Experimental and Theoretical Physics \textbf{107},
  668--678 (2008)

\bibitem{ZubarevKarabutJETPLett2018eng}
Zubarev, N.M., Karabut, E.A.: {Exact Local Solutions for the Formation of
  Singularities on the Free Surface of an Ideal Fluid}.
\newblock JETP Letters \textbf{107}, 412--417 (2018)

\bibitem{Zubarev_Kuznetsov_JETP_2014}
Zubarev, N.M., Kuznetsov, E.A.: {Singularity Formation on a Fluid
Interface
  During the Kelvin-Helmholtz Instability Development}.
\newblock J. Exp. Theor. Phys. \textbf{119}, 169--178 (2014)

\end{thebibliography}

%\begin{thebibliography}{}
%

%\end{thebibliography}

\end{document}